\documentclass[print,epsfig,onecolumn,floats,showpacs]{revtex4}
\usepackage{graphicx}
\usepackage{dcolumn}
\usepackage{bm}
\usepackage{amssymb}
\usepackage{amsmath}
\usepackage{epsfig}
\usepackage{amsmath}
\allowdisplaybreaks[4]
\begin{document}
\title{Generalized Remote Preparation of Arbitrary $m$-qubit Entangled States via Genuine Entanglements}
\author{Dong Wang$^{1,2,3,}$\footnote{{dwang@ahu.edu.cn (D. Wang)}}, Ross D. Hoehn $^{2,}$\footnote{{rhoehn@purdue.edu (R. Hoehn)}}, Liu Ye $^{1,}$\footnote{{yeliu@ahu.edu.cn (L. Ye)}} and Sabre Kais $^{2,4,}$\footnote{{kais@purdue.edu (S. Kais)}}}
\affiliation{${\ ^1}$ School of Physics \& Material Science, Anhui University, Hefei
230601, China
\\
${\ ^2}$ Department of Chemistry and Birck Nanotechnology Center,
Purdue University, West Lafayette, IN 47907, USA\\
${\ ^3}$
National Laboratory for Infrared Physics, Shanghai Institute of Technical Physics, Chinese Academy of Sciences, Shanghai 200083, China\\
${\ ^4}$Qatar Environment and Energy Research Institute, Qatar Foundation, Doha, Qatar
}

\begin{abstract}
Herein, we present a feasible, general protocol for quantum communication within a network via generalized remote preparation of an arbitrary $m$-qubit entangled state designed with genuine tripartite Greenberger--Horne--Zeilinger-type entangled resources. During the implementations, we construct novel collective unitary operations; these operations are tasked with performing the necessary phase transfers during remote state preparations. We have distilled our implementation methods into a five-step procedure, which can be used to faithfully recover the desired state during transfer. Compared to previous existing schemes, our methodology features a greatly increased success probability. After the consumption of auxiliary qubits and the performance of collective unitary operations, the probability of successful state transfer is increased four-fold and eight-fold for arbitrary two- and three-qubit entanglements when compared to other methods within the literature, respectively. We conclude this paper with a discussion of the presented scheme for state preparation, including: success probabilities, reducibility and generalizability.
\begin{flushleft}
\ \ \ \ \ \ \ \ \ \ \ \ \ \ \ \ \ \ \ \ \ \ \ \ \  operation; success probability
\end{flushleft}
\end{abstract}
\maketitle

\section{\label{sec:level1}Introduction}

Quantum entanglement is the primary resource for both quantum computation and quantum communication. Utilizing these resources allows one to perform information processing with unprecedented high efficiencies by exploiting the fundamental laws of quantum mechanics. Specifically, quantum entanglement possesses a variety of intriguing applications within the realm of quantum information processing \cite{C.H. Bennett,H.K.,A.K. Pati,C.H.,Mark H,Ryszard H,KaisSabre,Kais,ZhuJing}; these applications include: quantum teleportation (QT) \cite{C.H. Bennett}, remote state preparation (RSP) \cite{H.K.,A.K. Pati,C.H.}, quantum secret sharing \cite{Mark H}, quantum cryptography \cite{Ryszard H}, \textit{etc}. Both QT and RSP are important methods in quantum communication. With the help of previously-shared entanglements and necessary classical communications, QT and RSP can be applied to achieve the transportation of the information encoded by qubits. Yet, there exists several subtle differences between QT and RSP, including: classical resource consumptions and the trade-off between classical and quantum resources. Typically in standard QT, the transmission of an unknown quantum state consumes 1 ebit and an additional 2 cbits. In contrast, if the state is known to the sender, the resources required for the same action can be reduced to 1 ebit and 1 cbit in RSP. This decrease in resource consumption generally comes at the expense of a lower success probability. Furthermore, Pati \cite{A.K. Pati} has argued that RSP is able to maintain its low resource consumption while meeting the success probability of QT for preparing special ensemble states (e.g., states existing on the equator and great polar circle of the Bloch sphere). Characterized by conservation of resources while maintaining high total success probability (TSP), it is not surprising that RSP has recently received much attention within the literature.

To date, many authors have proposed a number of promising methodologies for RSP; a list of such methods should include: low-entanglement RSP \cite{Devetak}, optimal RSP \cite{Leung1}, oblivious RSP \cite{Berry,Kurucz1}, RSP without oblivious conditions \cite{Hayashi}, generalized RSP \cite{Abeyesinghe}, faithful RSP \cite{Ye}, joint RSP (JRSP) \cite{Xia1,Nguyen1,Luo1,Xiao-Qi Xiao22,Qing-Qin Chen4,Nguyen2,Luo2,Ping Zhou6, Yan Xia3,You-Ban Zhan6,Qing-Qin Chen5,Kui Hou,Ming Jiang2,Jia-Yin Peng4,Yue-Ming Liao2, Xiu-Bo Chen2,Zhi-Hua Zhang}, multiparty-controlled JRSP \cite{Dong1}, RSP for many-body states \cite{Yu,Huang,Liu1,Liu2,Liu3,Wang1,Dai1,Dai2,Yan} and continuous variable RSP in phase space \cite{Paris,Kurucz2}. Various RSP proposals utilizing different physical systems have been experimentally demonstrated, as well \cite{Peng,Xiang,Peters,Jeffrey,Mikami,Julio T. Barreiro,Magnus,Liu4,Wu1}. For example, Peng {\it et al.} investigated an RSP scheme employing NMR techniques \cite{Peng}, while others have explored the use of spontaneous parametric down-conversion within their RSP schemes \cite{Xiang, Peters}. Mikami {\it et al.} \cite{Mikami} experimentally demonstrated a novel preparation method for an arbitrary, pure single-qutrit state via biphoton polarization; furthermore, they claim that their method requires only two single-qubit projective measurements without any interferometric setup. Barreiro {\it et al.} \cite{Julio T. Barreiro} reported the remote preparation of two-qubit hybrid entangled states, including a family of vector-polarization beams; the single-photon states are encoded within the photon spin and orbital angular momentum, and the desired state is reconstructed by means of spin-orbit state tomography and transverse polarization tomography. Very recently, R{\aa}dmark {\it et al.} \cite{Magnus} experimentally demonstrated multi-location remote state preparation via multiphoton interferometry. This method allows the sender to remotely prepare a large class of symmetric states (including single-qubit states, two-qubit Bell states and three-qubit $W$, or $\overline{W}$ states).

There do exist a number of proposals \cite{Jin-M,Xiu7,SongM,You-B} dedicated to addressing the RSP of arbitrary two- and three-qubit entangled pure states. Liu {\it et al.} employed two and three Bell-type entanglements as quantum channels for conducting such preparations with total success probabilities (TSP) of $(a_1a_2)^2$ and $(a_1a_2a_3)^2$, respectively \cite{Jin-M}. Both Brown and $\chi$ states have also been employed for the creation of correlations among participants \cite{Xiu7,SongM}. Resulting from these correlations, the maximal success probability for general two- and three-qubit states is $\leq 50\%$ for such strategies. Recently, Zhan \cite{You-B} presented two schemes for the remote preparation of two- and three-qubit entangled states with unity success probability via maximally entangled states, \textit{i.e}., Greenberger--Horne--Zeilinger (GHZ) states. In our present work, the aim is to investigate generalized remote preparation for an arbitrary $m$-qubit entangled state, while only utilizing general entanglement states (\textit{i.e}., non-maximally entangled states) as quantum channels. We will show that the above scheme is capable of performing faithful RSP with a four-fold or eight-fold increase of the success probability over existing methods, for $m=2$ and $m=3$, respectively \cite{Jin-M}. These enhancements are afforded by the construction of two novel $m$-qubit collective unitary transformations, respective of the number of entangled qubits within the desired state.

The organization of this paper is as follows: In the next section, we shall detail our procedure for the RSP of a general $m$-qubit entangled state employing a series of GHZ-type entanglements as quantum channels. Our results show that the desired state can be faithfully restored at the receiver with a fixed, predictable success probability. In Section \ref{section:3}, we will illustrate our general procedure through its implementation for the RSP of a two-qubit entangled state. Section \ref{section:4} will contain our discussion and comments on the procedure, as well as an evaluation of the classical information cost (CIC) for the procedure and the total success probability (TSP), which can be expected. We will close with Section~\ref{section:5}, containing a concise summary. We have also chosen to attach a second illustrative example in the {Appendix}; this example repeats the general procedure upon a three-qubit entangled state.

\section{General RSP Procedure for an Arbitrary $m$-Qubit Entangled State}\label{section:2}

The method presented in this paper is a general scheme for the remote preparation of an arbitrary state using a generic ($m$) number of GHZ-type entanglements, which will be used as quantum channels. Within this procedure, we will firstly specify an $m$-qubit state, which we desire to be transferred from a sender (Alice) to a receiver (Bob). For simplicity, we have introduced $\mathit{l}=2^{m}$ to note the number of vectors required to form a complete basis set for a set of $m$ qubits. Furthermore, we introduce $\mathit{n}=3m$ as the number of qubits required to form the GHZ-entangled quantum channels. The desired state is given~by:
\begin{equation}
|{\cal{P}}\rangle = \alpha_0 |00 \ldots 0\rangle + \alpha_1 e^{i\eta_1} |00 \ldots 1\rangle + \ldots + \alpha_{ \mathit{l} -1} e^{i\eta_{ \mathit{l} -1}} |11 \ldots 1\rangle.
\end{equation}
Within the above, constraints are imposed on the coefficients and phase factors: $\alpha_{i}\in\mathbb{R}$ and satisfies the normalized condition $\sum_{i=0}^{\mathit{l}-1}{\alpha_{i}}^2=1$; and $\eta_{i}\in[0,2\pi]$. The series of GHZ-type entangled states used as quantum channels are given by:
\begin{eqnarray}
|&\phi_{1}&\rangle = ( x_0 |000\rangle + y_0 |111\rangle )_{123}; \\
|&\phi_{2}&\rangle = ( x_1 |000\rangle + y_1 |111\rangle )_{456}; \\
&\vdots & \\
|&\phi_{m}&\rangle = ( x_{m-1} |000\rangle + y_{m-1} |111\rangle )_{( \mathit{n} -2)( \mathit{n} -1)( \mathit{n} )}.
\end{eqnarray}
Without loss of generality, we may assert the following two constraints: $x_{i} \in \mathbb{R}$ and $|x_i|\leq|{y}_i|$. Within these channels, a series of qubits are held locally by Alice, $ \{ 1, 2, 4, 5, \ldots, \mathit{n} -2, \mathit{n} -1 \} $, and another by Bob, $\{ 3, 6, \ldots, \mathit{n} \}$. We can now proceed to our stepwise procedure for RSP.

{\bf Step 1:} Alice will perform an $m$-partite projective measurement on qubits: $(1, 4, \ldots, \mathit{n}-2)$. This measurement is defined through application of a projection matrix, $\Omega$, which is constructed by starting with a state that is similar to the desired state, $|{\cal{P}}\rangle$, except opposite phases, as the first row vector and producing a series of orthonormal spanning vectors. The aforementioned series of qubits is now described by a new complete series of orthogonal vectors: $\{|{M}_{i_{1}i_{2} \ldots i_{m} }\rangle_{14 \ldots \mathit{n} -2 }\}$,
where $ \{ i_1, i_2, \ldots, i_{ m } \} \in \{ 0, 1 \}$. This series of spanning vectors is comprised of orthonormal weightings of the $\mathit{l}$-dimensional ordering basis.

The resulting $\mathit{n}$-qubit systemic state, $ | \Phi \rangle $, taken as quantum channels is factorizable as:
\begin{equation}
\begin{split}
| \Phi \rangle &= |\phi_{1} \rangle \otimes |\phi_{2}\rangle \otimes \ldots \otimes |\phi_{m}\rangle \\
&= \sum_{i_1, i_2, \ldots i_m}^{0, 1 } |{M}_{i_{1}i_{2} \ldots i_{m} }\rangle_{14 \ldots \mathit{n} -2 } \otimes |{R}_{i_{1}i_{2} \ldots i_{m} }\rangle_{2356 \ldots ( \mathit{n} -1)( \mathit{n}) } .
\end{split}
\end{equation}
where in the above, the non-normalized state $|{R}_{i_{1}i_{2} \ldots i_{\mathit{m}} }\rangle_{2356 \ldots ( \mathit{n} -1)( \mathit{n}) } \equiv \,_{14 \ldots \mathit{n} -2 }\langle {M}_{i_{1}i_{2} \ldots i_{\mathit{m}} } | \Phi \rangle $ can be probed with specific probability, $ ( 1 / N_{i_1 i_2 \ldots i_m} )^{2} $, where $ N_{i_1 i_2 \ldots i_m}$ is the normalization coefficient of state $| {R}_{ i_{1} i_{2} \ldots i_{ m } } \rangle_{ 2 3 5 6 \ldots ( \mathit{n} -1 ) \, (\mathit{n}) } $.

{\bf Step 2:} Alice executes a second $m$-partite joint unitary operation constructed under the $\mathit{l}$-dimensional ordering basis. This operation will be designated by the operator form: $\hat{\cal{U}}_{2 5 \ldots ( \mathit{n}-1 ) }^{i_1 i_2 \ldots i_{m}}$. She does this operation on qubits $(2, 5, \ldots, \mathit{n}-1)$. This operation is designed to canonically order the phase factors, $\eta_i$, within the set of $ \{ |{R}_{i_{1} i_{2} \ldots i_{\mathit{m}} }\rangle_{2 3 5 6 \ldots ( \mathit{n}-1 ) ( \mathit{n} ) }\}$ vectors. There will be $\mathit{l}$ unique operators of this class; Alice selects the operator in response to the outcome of the previous measurement.

{\bf Step 3:} Alice now measures qubits $(2, 5, \ldots, \mathit{n}-1)$ under the complete set of orthogonal basis vectors: $\{ | \pm \rangle \}$. She then has%please double check the correction
 all of her measurement outcomes via classical channels. To conserve the amount of classical resources required for this system, it should be prearranged that all authorized anticipators of this information that are%please double check the correction
 cbits $(i_1 i_2 \dots{i_m})$ will correspond to the outcome of the measurement in Step 1 and cbits $(j_1 j_2 \ldots{j_m})$ will designate the outcomes of the measurement in Step~3. For brevity, we shall declare that the authorized anticipators have been prearranged to use the cbit notation:
$$j_k (k=1,2,\ldots,m)=\left\{
\begin{array}{rcl}
0, & & \ {\rm if}\ {|+\rangle}\ {\rm is} \ {\rm measured}\\
1, & & \ {\rm if}\ {|-\rangle}\ {\rm is} \ {\rm measured}
\end{array} . \right . $$

Before moving on to Step 4, we should first comment on a special case. In the limiting case where we take the quantum channels to be maximally entangled, we may omit Step 4 and move to Step 5. Only in the case where the channels are permitted to assume an arbitrary degree of entanglement is Step 4 necessary.

{\bf Step 4:} Bob now introduces a single auxiliary qubit, $A$, in an initial state $ | 0 \rangle$. He then performs a $(m+1)$-partite collective unitary transformation, $\hat{\cal{U}}_{3 6 \ldots (\mathit{n}) A}$, on qubits $(3, 6, \ldots, \mathit{n}, A)$ under the $2^{m+1}$ series of ordering basis vectors. This operator takes the form of a $2^{m+1} \times 2^{m+1}$ matrix, whose intent is the resolution of the $x_i$ and $y_i$ coefficients from the state vectors.

Since Bob does not possess knowledge of the desired state, $ | \cal{P} \rangle$, he is unable to ascertain the success of the protocol. For this reason, Bob then measures qubit $A$ under the measuring basis vectors $\{ | 0 \rangle, | 1 \rangle \}$. If Bob detects state $ | 1 \rangle $, the remaining qubits at his location will collapse into the trivial state, which will be the fail state for the RSP procedure; he must start from the beginning. If state $ | 0 \rangle $ is detected, the procedure may continue forth to the next step.

{\bf Step 5:} Bob, aware of the success of the RSP transfer, now is able to reclaim the desired state, $ | \cal{P} \rangle$. Then there is the%please double check the addition
 application of a final operation upon the qubits at Bob's location (qubits $\{ 3, 6, \ldots, \mathit{n} \}$). This operation is specifically tuned to information contained within the classical bit sequence that Bob received from Alice $( i_1 i_2 \ldots i_{m} j_1 j_2 \ldots j_{m} )$ and is denoted by $\hat{\cal{U}}_{3 6 \ldots \mathit{n}}^{ i_1 i_2 \ldots i_{m} j_1 j_2 \ldots j_{m} }$.

\section{RSP for an Arbitrary Two-Qubit Entangled State: An Example}\label{section:3}

Here, we shall illustrate the above procedure through an example. We have selected a small value for our m%italics or not? please check the convention for math terms throughout
-qubit entangled state, $m = 2$. Furthermore, for the benefit of comparing and contrasting, we have included a second example, $m=3$, as the Appendix to this paper. Let us now define the desired state that Alice wishes to prepare in Bob's distant laboratory. The desired state, $|{\cal{P}}\rangle$, is an arbitrary two-qubit entangled state given by:
\begin{equation}
|{\cal{P}}\rangle=\alpha_0|00\rangle+\alpha_1e^{i\eta_1}|01\rangle+\alpha_2e^{i\eta_2}|10\rangle+
\alpha_3e^{i\eta_3}|11\rangle.
\end{equation}
where $\alpha_{i}\in\mathbb{R}$ and satisfies the normalized condition
$\sum_{i=0}^3{\alpha_{i}}^2=1$; furthermore, $\eta_{i}\in[0,2\pi]$. Note that,
Alice has knowledge of the desired state, yet Bob has no such knowledge.
Initially, a class of robust and genuine GHZ-type entanglements must be constructed and shared between Alice and Bob. These GHZ states for our example are given by:
\begin{equation}
|\phi_{1}\rangle=(x_0|000\rangle+y_0|111\rangle)_{123},
\end{equation}
and:
\begin{equation}
|\phi_{2}\rangle=(x_1|000\rangle+y_1|111\rangle)_{456}.
\end{equation}
Without loss of generality, the conditions ${x}_i\in\mathbb{R}$ and $|x_i|\leq|{y}_i|$ are maintained. Initially, Qubits 1, 2, 4 and 5 are held by Alice, while Qubits 3 and 6 are held by Bob.

In order to accomplish our RSP procedure, we shall implement the steps discussed within Section \ref{section:2}:

{\bf Step 1:}
Alice executes one bipartite projective measurement, $\Omega $, on the qubit bipartite ($1,4$)
under a set of complete orthogonal basis vectors $\{|{M}_{ij}\rangle_{14}\}$, where the indices $i,j\in\{0,1\}$ take the place of $(i_1, i_2, \ldots)$ within the general procedure. This basis, $\{|{M}_{ij}\rangle_{14}\}$, is written in terms of the computational basis:
$\{|00\rangle,|01\rangle,|10\rangle,|11\rangle\}$. The projective measurement is formed in the method previously discussed and can be written as:
\begin{eqnarray}
{\Omega}=\left(\begin{array}{cccc}
\alpha_0 & \alpha_1{e^{-i\eta_1}} & \alpha_2{e^{-i\eta_2}} & \alpha_3{e^{-i\eta_3}} \\
\alpha_1 & -\alpha_0{e^{-i\eta_1}} & \alpha_3{e^{-i\eta_2}} & -\alpha_2{e^{-i\eta_3}} \\
\alpha_2 & -\alpha_3{e^{-i\eta_1}} & -\alpha_0{e^{-i\eta_2}} & \alpha_1{e^{-i\eta_3}} \\
\alpha_3 & \alpha_2{e^{-i\eta_1}} & -\alpha_1{e^{-i\eta_2}} & -\alpha_0{e^{-i\eta_3}} \\
\end{array}\right).
\end{eqnarray}
The result of the projective transformation is:
\begin{eqnarray}
\left(|M_{00}\rangle_{14}, |M_{01}\rangle_{14}, |M_{10}\rangle_{14}, |M_{11}\rangle_{14}
\right)^T ={\Omega}\cdot\left(|00\rangle, |01\rangle ,|10\rangle ,|11\rangle \right)^T.
\end{eqnarray}

As a result, our quantum channels, constructed from the six-qubit systemic state, can be expressed~as:
\begin{equation}
\begin{split}
|\Phi\rangle&=|{\phi}_{1}\rangle_{123}\otimes|\phi_{2}\rangle_{456}\\
&=\sum_{i,j}^{0,1}|{M}_{ij}\rangle_{14}\otimes|{R}_{ij}\rangle_{2356}\\
&=
|{M}_{00}\rangle_{14}(\alpha_0x_0x_1|0000\rangle+\alpha_1e^{i\eta_1}x_0y_1|0011\rangle
+\alpha_2e^{i\eta_2}y_0x_1|1100\rangle+\alpha_3e^{i\eta_3}y_0y_1|1111\rangle)_{2356}\\
&+|{M}_{01}\rangle_{14}(\alpha_1x_0x_1|0000\rangle-\alpha_0e^{i\eta_1}x_0y_1|0011\rangle
+\alpha_3e^{i\eta_2}y_0x_1|1100\rangle-\alpha_2e^{i\eta_3}y_0y_1|1111\rangle)_{2356}\\
&+|{M}_{10}\rangle_{14}(\alpha_2x_0x_1|0000\rangle-\alpha_3e^{i\eta_1}x_0y_1|0011\rangle
-\alpha_0e^{i\eta_2}y_0x_1|1100\rangle+\alpha_1e^{i\eta_3}y_0y_1|1111\rangle)_{2356}\\
&+|{M}_{11}\rangle_{14}(\alpha_3x_0x_1|0000\rangle+\alpha_2e^{i\eta_1}x_0y_1|0011\rangle -\alpha_1e^{i\eta_2}y_0x_1|1100\rangle-\alpha_0e^{i\eta_3}y_0y_1|1111\rangle)_{2356}.
\end{split}
\end{equation}
where the non-normalized state $|{R}_{ij}\rangle_{2356}\equiv{_{14}}\langle{M}_{ij}|\Phi\rangle$ can be probed with the probability $ ( 1 / N_{ij} ) ^{2}$.

{\bf Step 2:}
Following the measurement $|{M}_{ij}\rangle$, Alice executes a corresponding bipartite joint unitary operation, $\hat{\cal U}_{25}^{ij}$, on Qubits 2 and 5, under the ordering basis:
$\{|00\rangle,|01\rangle,|10\rangle,|11\rangle\}$. To be explicit, $\hat{\cal U}_{25}^{ij}$ is taken as a $4\times4$ matrix of the form:
\begin{eqnarray}
\hat{\cal{U}}_{25}^{00}&=&{\rm diag}(1, 1, 1, 1)=\mathbb{I}_{4\times4},\\
\hat{\cal{U}}_{25}^{01}&=&{\rm diag}(
e^{i\eta_1}, -e^{-i\eta_1}, e^{i(\eta_3-\eta_2)}, -e^{i(\eta_2-\eta_3)}),\\
\hat{\cal{U}}_{25}^{10}&=&{\rm diag}(
e^{i\eta_2}, -e^{i(\eta_3-\eta_1)}, -e^{-i\eta_2}, e^{i(\eta_1-\eta_3)}),\\
\hat{\cal{U}}_{25}^{11}&=&{\rm diag}(
e^{i\eta_3}, e^{i(\eta_2-\eta_1)}, -e^{i(\eta_1-\eta_2)}, -e^{-i\eta_3}).
\end{eqnarray}

{\bf Step 3:} Alice now performs a measurement on Qubits 2 and 5 under the complete set of orthogonal basis vectors: $\{|\pm\rangle:=\frac1{\sqrt2}(|0\rangle\pm|1\rangle)\}$. She then publishes the measurement outcomes via classical channels where the authorized anticipators have already conspired concerning the interpretation of the classical bits. It should now be stated that all authorized anticipators have conspired in advance that cbits $'ij'$ correspond to the outcome
$|{M}_{ij}\rangle_{14}$ and cbits $'rs'$ (previously $(j_1, j_2, \ldots)$) correspond to the measuring outcome of Qubits 2 and 5.

{\bf Step 4:} Bob introduces one auxiliary qubit, $A$, with an initial state of $|0\rangle$. He then performs a triplet collective unitary transformation, $\hat{\cal U}_{36A}$, on Qubits 3, 6 and $A$ under the set of ordering basis vectors: $\{|000\rangle_{36A},$ $|010\rangle_{36A},$ $|100\rangle_{36A},$ $|110\rangle_{36A},$ $|001\rangle_{36A},$ $|011\rangle_{36A},$ $|101\rangle_{36A}$, $|111\rangle_{36A}\}$. The transformation matrix is given by:
\begin{eqnarray}
\hat{\cal U}_{36A}=\left(
\begin{array}{cc}
{\hat{\cal D}} & {\hat{\cal F}}\\
{\hat{\cal F}} & -{\hat{\cal D}}\\
\end{array}
\right)_{8\times8},
\end{eqnarray}
where the operators ${\hat{\cal{D}}}$ and ${\hat{\cal{F}}}$ are $4\times4$ matrices. Explicitly, these operators are given by:
\begin{eqnarray}
{\hat{\cal{D}}}={\rm diag}(
1,\frac{x_1}{y_1},\frac{x_0}{y_0},\frac{x_0x_1}{y_0y_1})
\end{eqnarray}
and:
\begin{equation}\begin{split}
{\hat{\cal{F}}}={\rm diag}(
0,\sqrt{1-(\frac{x_1}{y_1})^2},\sqrt{1-(\frac{x_0}{y_0})^2},\sqrt{1-(\frac{x_0x_1}{y_0y_1})^2}).
\end{split}\end{equation}

Subsequently, Bob measures qubit $A$ under a set of measuring basis vectors,
$\{|0\rangle,|1\rangle\}$. If state $|1\rangle$ is detected, his remaining
qubits will collapse into the trivial state. If
$|0\rangle$ is obtained, the preparation procedure may continue on to the final step.

{\bf Step 5:} Finally, Bob executes an appropriate unitary transformation, $\hat{{\cal {U}}}_{36}^{ijrs}$ (see Table 1 for more details), on his Qubits 3 and 6. The exact form of this operator varies with the observed values associated with the measurements denoted by cbits $i$, $j$, $r$ and $s$. This operation allows Bob to recover $ | \cal{P} \rangle$ at his location.

\begin{figure}
\centering
\includegraphics[scale=1]{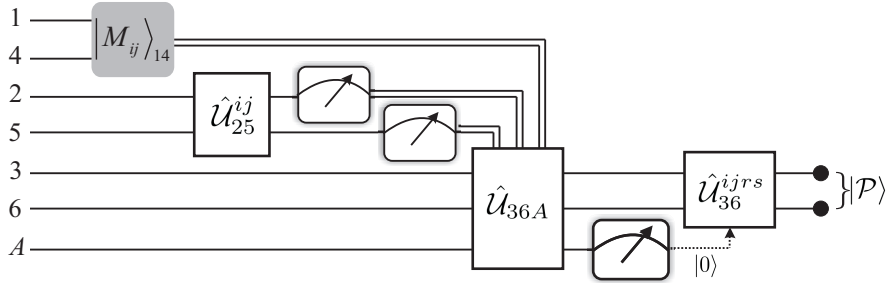}
\caption{Quantum circuit for implementing remote state preparation (RSP) of arbitrary two-qubit entangled states. $|{M}_{ij}\rangle_{14}$ denotes a two-qubit projective measurement on
Qubits~1 and 4 under a set of complete orthogonal basis vectors $\{|{M}_{ij}\rangle_{14}\}$; $\hat{\cal{U}}_{25}^{ij}$ denotes Alice's appropriate collective unitary transformation on bipartite (2,5); $\hat{\cal{U}}_{36A}$ denotes Bob's collective three-qubit unitary transformation on his Qubits 3, 6 and $A$, and $\hat{\cal{U}}_{36}^{ijrs}$ denotes Bob's appropriate single-qubit unitary transformations on his Qubits 3 and 6.}
\label{fig:1}
\end{figure}

This overall procedure may be conveyed as a quantum circuit and is displayed within Figure \ref{fig:1}.

\begin{table}
\centering
\begin{tabular}{|l|l|l|l|l|l|l|l|}
\hline $ijrs$ & $\hat{\cal{U}}_{36}^{ijrs}$ & $ijrs$ & $\hat{\cal{U}}_{36}^{ijrs}$ & $ijrs$ &
$\hat{\cal{U}}_{36}^{ijrs}$& $ijrs$ & $\hat{\cal{U}}_{36}^{ijrs}$ \\ \hline
$0000$ & $I_{3}I_{6}$ & $0100$ & $I_{3}\sigma^x_{6}$ & $1000$ & $\sigma^x_{3}I_{6}$ &$1100$ & $\sigma^x_3\sigma^x_6$ \\
$0001$ & $I_{3}\sigma^z_{6}$ & $0101$ & $I_{3}\sigma^x_{6}\sigma^z_{6}$ & $1001$ & $\sigma^x_{3}\sigma^z_{6}$ &$1101$ & $\sigma^x_3\sigma^x_6\sigma^z_6$ \\
$0010$ & $\sigma^z_{3}I_{6}$ & $0110$ & $\sigma^z_{3}\sigma^x_{6}$ & $1010$ & $\sigma^x_{3}\sigma^z_3I_{6}$ &$1110$ & $\sigma^x_3\sigma^z_3\sigma^x_6$ \\
$0011$ & $\sigma^z_{3}\sigma^z_{6}$ & $0111$ & $\sigma^z_{3}\sigma^x_{6}\sigma^z_{6}$ & $1011$ & $\sigma^x_{3}\sigma^z_3\sigma^z_{6}$ &$1111$ & $\sigma^x_3\sigma^z_3\sigma^x_6\sigma^z_6$ \\
\hline
\end{tabular}
\caption{$ijrs$ denotes the series of cbits corresponding to measurement outcomes from the sender and $\hat{\cal{U}}_{36}^{ijrs}$
denotes an unitary transformation that Bob needs to perform on Qubits 3 and 6 for the recovery \mbox{of $| \cal{P} \rangle$}.}

\end{table}

\section{Discussion}\label{section:4}
\vspace{-12pt}
\subsection{Total Success Probability and Classical Information Cost}

\begin{figure}
\includegraphics[width=\textwidth]{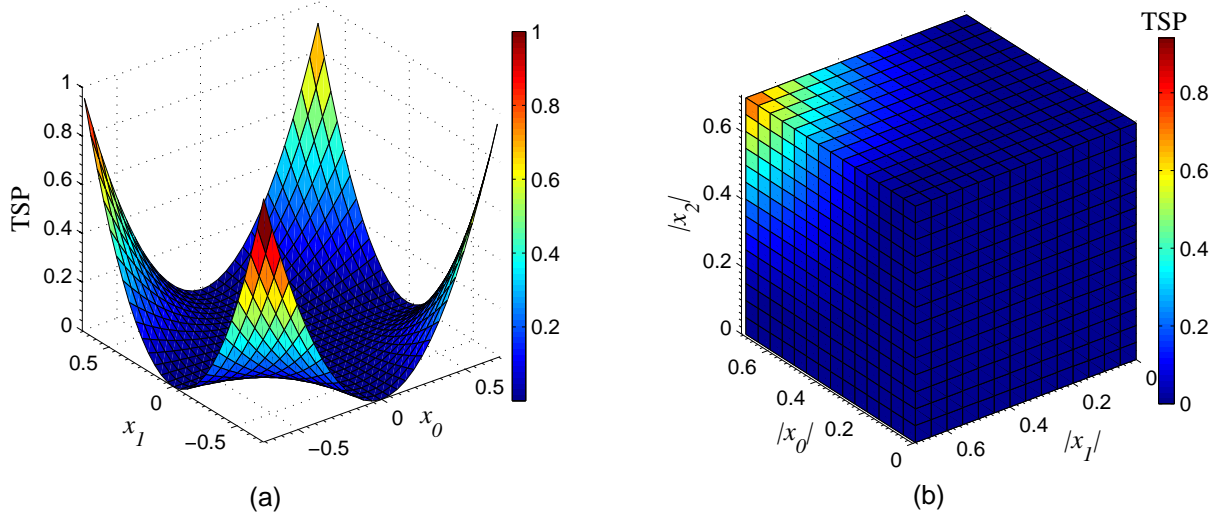}
\caption{The relation between the total success probability (TSP) and the smaller coefficients of entanglements severing as quantum channels. (\textbf{a}) The case of RSP for
arbitrary two-qubit entangled states; (\textbf{b}) the case of RSP for
arbitrary three-qubit entangled states. One can see that the TSP is increased as the value of $|x_i|$ increases.}
\label{fig.3}
\end{figure}

In this subsection, let us turn to calculate the TSP and CIC of the present scheme. In our generalized scheme, one can see from the discussion of Step 1 in Section \ref{section:2} that the state ${|{M}_{i_1i_2\ldots{i_m}}\rangle}$ can be probed with probability of:
\begin{eqnarray}
P({|{M}_{i_1i_2\ldots{i_m}}\rangle})=\frac1{(N_{i_1i_2\ldots{i_m}})^2} .
\end{eqnarray}
The probability for the capture of $|0\rangle_{A}$ is given by:
\begin{eqnarray}
P({|0\rangle_{A}})={({N}_{i_1i_2\ldots{i_m}} x_0 x_1 \ldots { x_{m-1}})}^2.
\end{eqnarray}
Hence, the success probability of RSP for the particular measurement outcome $(i_1i_2\ldots{i_m})$ is equal to:
\begin{eqnarray}
P{(i_1i_2\ldots{i_m})}=P({|{\cal A}_{ij}\rangle})\times{P}({|0\rangle_{A}})=( x_0 x_1 \ldots { x_{m-1}})^2.
\end{eqnarray}
It can easily be determined that the TSP over all possible states sums to:
\begin{eqnarray}
P_{Total}=\sum_{i_1,i_2,\ldots,{i_m}}^{0,1}{P}{(\alpha_0\alpha_1\ldots{\alpha_{m-1}})}=2^m(x_0 x_1 \ldots {x_{m-1}})^2 ,
\end{eqnarray}
which is inherently associated with the smaller coefficients of the employed channels. The interplay between the choice of coefficients and the TSP can most easily be seen in Figure \ref{fig.3} for the $m=2$ and $m=3$ example systems.

Moreover, one can work out that the required CIC should be of the form:
\begin{eqnarray}
C(P)=2^{m+1}( x_0 x_1 \ldots { x_{m-1}})^2{\log}_2\frac1{( x_0 x_1 \ldots { x_{m-1}})^2}\ ({\rm cbits}) .
\end{eqnarray}
This value, $ C(P)$, is constructed as an {average} relying on the definition of resource consumption in \cite{Dai2} and necessary extra communication for outcome $(j_1,j_2,\ldots,j_m)$ between the sender and the receiver.

\subsection{The Properties of the Current Scheme}

We have also found that there are several remarkable properties with respect to our present scheme; these include: (1) {high success probability}.
Generally, contemporary RSP protocols can be faithfully performed with a TSP of $2^m( x_0 x_1 \ldots { x_{m-1}})^2$.
When $| x_i|=1/\sqrt2$ is chosen, the TSP can be pushed as high as one. (2) {Reducibility.} Within our scheme, if $m$ is reduced to two and three, two specific schemes naturally appear: RSP for arbitrary two- and three-qubit entangled states. Furthermore, with respect to RSP for two- and three-qubit states, some applicable schemes have already been presented, which permits a degree of comparison \cite{Jin-M,Xiu7,SongM,You-B}. There do, however, exist differences in key elements associated with intrinsic efficiency, including operation complexity and resources consumption. We have provided a comparison between RSP schemes for maximally-entangled states as Table \ref{tab.2}, illustrated by items, such as: required quantum resource, the necessary classical resource, the required operation, success probability and intrinsic efficiency. We should also stress that in the limit where our quantum channels are taken to be maximally entangled, Step 4 become needless. Stated otherwise, Steps 1--3 and 5 are sufficient to achieve RSP for an arbitrary $m$-qubit state when we restrict our channels to be maximally entangled.

\begin{table*}
\caption{Comparison between our scheme and the previous works in the case of
maximally entangled channels.
Within this table abbreviations should be read as: {EPR : Einstein-Podolsky-Rosen entangled state}; {BS : Brown state}; {GHZ : Greenberg-Horne-Zeilinger state}; {TQPM : two-qubit projective measurement};
{SQPM : single-qubit projective measurement};
{TEQPM : three-qubit projective measurement}; {FQPM : four-qubit projective measurement}; {CIC : classical information consumption}; {TSP : total success probability}; and $\Gamma$ represents intrinsic efficiency
of the scheme.}
\begin{center}
\label{tab.2}
\begin{tabular}{cccccccc}
\hline Protocols & Entanglements employed & Quantum operations & CIC & TSP & $\Gamma$\\ \hline
Two-qubit case \\
Ref. \cite{Jin-M} & two 2-qubit EPR & one TQPM & 2 & $\frac1{4}$ & 8.33\% \\
Ref. \cite{Xiu7} & five-qubit BS & one TQPM \& one SQPM & 3 & $\frac12$ & 12.5\% \\
Ref. \cite{SongM} & five-qubit $\chi$-state & one TEQPM & 3 & $\frac12$ & 12.5\% \\
Our scheme & two 3-qubit GHZ & one TQPM \& two SQPM & 4 & 1 & 20\% \\ \hline
Three-qubit case \\
Ref. \cite{Jin-M} & three 2-qubit EPR & one TEQPM & 3 & $\frac1{8}$ & 8.33\% \\
Ref. \cite{Xiu7} & five-qubit BS \& EPR & one TEQPM \& one SQPM & 4 & $\frac12$ & 13.64\% \\
Ref. \cite{SongM} & four-qubit $\chi$-state \& GHZ & one FQPM & 4 & $\frac12$ & 13.64\% \\
Our scheme & three GHZ & one TEQPM \& two SQPM & 6 & 1 & 20\% \\
\hline
\label{tab.2}
\end{tabular}
\end{center}
\end{table*}

From Table \ref{tab.2}, it can be directly noted that the TSP of our scheme is capable of both approaching and attaining a value of unity. The intrinsic efficiency $(\Gamma)$ achieves $20\%$, which is much greater than that of previous schemes \cite{Jin-M,Xiu7,SongM}. Due to characteristically high intrinsic efficiency and TSP, our scheme is highly efficient when compared to other existent schemes; further, our scheme is capable of optimal performance in specific limiting cases.
\textit{En passant}, the intrinsic efficiency of a scheme is defined by \cite{Cabello A.} and is given by the form:
\begin{equation}
\Gamma=\frac{Q_s}{Q_q+Q_c}\times{TSP}.
\end{equation}
In the above, $Q_s$ denotes the number of qubits in the desired states; $Q_q$ denotes the amount of quantum resources consumed in the process and $Q_c$ denotes the amount of classical information resources consumed. (3) {Generalizability}. Herein, we have designed a general scenario for RSP of arbitrary $m$-qubit states via GHZ-class entanglements. The generalization is embodied in several aspects, which we will now note. First, the states that we desire to remotely prepare are arbitrary $m$-qubit ($m=1,2,\ldots$) entangled states. Second, the quantum channels employed are GHZ-class entanglements, which are non-maximally-entangled states. It has been previously shown that non-maximally-entangled states are general cases and are more achievable in real-world laboratory conditions. In contrast, maximally-entangled states are a special case of general entangled states when the state coefficients are restricted to special values. Therefore, our scheme is a readily general procedure. Additionally, \cite{You-B} investigated deterministic RSP for both the $m=2$ and the $m=3$ cases; however, there are some differences between these schemes and the analogous cases within our works: First, \cite{You-B} concentrated only on the cases when maximally entangled states are taken as channels; this limit is just a special case of our schemes where the channels are general, yet still allow for the maximally-entangled case. Second, we employ a von Neumann projective measurement in a set of vectors $|\pm\rangle$ instead of measurement on the basis of $\{|0\rangle,|1\rangle\}$ and Hadamard transformations. Considering these differences, we argue that our scheme is more general than previous works, and we reduce both the number and complexity of operations in the overall procedure.

\section{Summary}\label{section:5}

In summary, we have derived a novel strategy for the implementation of RSP of a general $m$-qubit entangled state. This was done by taking advantage of robust GHZ-type states acting as quantum channels. With the assistance of appropriate local operations and classical communication, the schemes can be realized with high success probabilities, increased four-fold and eight-fold when compared to previous schemes with $m=2$ and $m=3$, respectively \cite{Jin-M}. Remarkably, our schemes feature several nontrivial properties, including a high success probability, reducibility and generalizability. Moreover, the TSP of RSP can reach unity when the quantum channels are reduced to maximally-entangled states; that is, our schemes become deterministic at such a limit. Further, we argue that our current RSP proposal might be important for applications in long-distance quantum communication using prospective node-node quantum networks.

\appendix
\vspace{12pt}
\noindent\textbf{Appendix}\\

Within this Appendix, we shall provide a second illustration of the procedure featured in Section \ref{section:2} of the main text. We have provided this example to assist in comparisons between two values of $m$ for the RSP of an arbitrary $m$-qubit entangled state. Appendix \ref{General-1} will cover the general RSP procedure for a three-qubit entangled state. Appendix \ref{General-2} will declare specific states for the measurements and explicitly perform the operations of the general procedure; $(i,j,k,r,s,t)=(0,0,1,0,0,1)$.

\section{General RSP for Three-Qubit Entangled States}
\label{General-1}

Let us attempt the RSP for an arbitrary three-qubit entangled state described by:
\begin{equation}
\begin{split}
| {\cal{P}} \rangle =& \alpha_0 |000\rangle + \alpha _1 e^ {i \eta_1 } |001\rangle + \alpha_2 e^{i\eta_2} |010\rangle +
\alpha_3 e^{i\eta_3} |011\rangle + \alpha_4 e^{i\eta_4} |100\rangle \\
&+ \alpha_5 e^{i\eta_5} |101\rangle + \alpha_6 e^{i\eta_6} |110\rangle + \alpha_7 e^{i\eta_7} |111\rangle .
\end{split}
\end{equation}
This state is to be remotely prepared at Bob's location, transmitted from Alice; in the above, the coefficients must satisfy the following conditions: $\alpha_{i}\in\mathbb{R}$, $\eta_{i}\in[0,2\pi]$ and $\sum_{i=0}^7{\alpha_i}^2=1$. It merits stressing that a nontrivial precondition in standard RSP must be met: the sender has the knowledge of the desired state, yet the receiver does not possess this knowledge. Originally, Alice and Bob are robustly linked by genuine entanglements (GHZ-type entanglements) described by:
\begin{equation}
|\phi_{1}\rangle=(x_0|000\rangle+y_0|111\rangle)_{123},
\end{equation}
\begin{equation}
|\phi_{2}\rangle=(x_1|000\rangle+y_1|111\rangle)_{456},
\end{equation}
and:
\begin{equation}
|\phi_{3}\rangle=(x_2|000\rangle+y_2|111\rangle)_{789}.
\end{equation}
We assume that the conditions ${x}_i\in\mathbb{R}$ and $|x_i|\leq|{y}_i|$ are satisfied.
Additionally, it should be noted that Qubits 1, 2, 4, 5, 7 and 8 are held by Alice, while Qubits 3, 6 and 9 are held by Bob.

For the sake of a successful RSP, the procedure can be implemented in a manner consistent with the five-step procedure in the main text:

{\bf Step 1:}
Alice performs a three-qubit projective measurement on the qubit triplet ($1,4,7$)
under a set of complete orthogonal basis vectors: $\{|{M}_{ijk}\rangle_{147}\}$ $(i,j,k\in\{0,1\})$; where $\{|{M}_{ijk}\rangle_{147}\}$ is comprised of this computational basis:
$\{|000\rangle$, $|001\rangle$, $|010\rangle$, $|011\rangle$, $|100\rangle$, $|101\rangle$, $|110\rangle$, $|111\rangle\}$. This measurement takes the form:
\begin{eqnarray}
\begin{array}{c}
\left(|M_{000}\rangle_{147}, \! |M_{001}\rangle_{147}, \! |M_{010}\rangle_{147}, \! |M_{011}\rangle_{147}, \! |M_{100}\rangle_{147}, \!
|M_{101}\rangle_{147}, \! |M_{110}\rangle_{147}, \! |M_{111}\rangle_{147} \! \right)^T\\
={\Omega}\cdot\left(|000\rangle, |001\rangle ,|010\rangle ,|011\rangle ,|100\rangle ,|101\rangle ,|110\rangle ,|111\rangle \right)^T.
\end{array}
\end{eqnarray}
where the projection operator, $\Omega$, is of the form:
\begin{eqnarray}
{\Omega} \! = \! \! \left( \! \! \! \begin{array}{cccccccc}
\alpha_0 & \alpha_1{e^{-i\eta_1}} & \alpha_2{e^{-i\eta_2}} & \alpha_3{e^{-i\eta_3}} &
\alpha_4{e^{-i\eta_4}}& \alpha_5{e^{-i\eta_5}}& \alpha_6{e^{-i\eta_6}}& \alpha_7{e^{-i\eta_7}}\\
\alpha_1 & -\alpha_0{e^{-i\eta_1}} & \alpha_3{e^{-i\eta_2}} & -\alpha_2{e^{-i\eta_3}} &
\alpha_5{e^{-i\eta_4}}& -\alpha_4{e^{-i\eta_5}}& \alpha_7{e^{-i\eta_6}}& -\alpha_6{e^{-i\eta_7}}\\
\alpha_2 & -\alpha_3{e^{-i\eta_1}} & -\alpha_0{e^{-i\eta_2}} & \alpha_1{e^{-i\eta_3}} &
-\alpha_6{e^{-i\eta_4}}& \alpha_7{e^{-i\eta_5}}& \alpha_4{e^{-i\eta_6}}& -\alpha_5{e^{-i\eta_7}}\\
\alpha_3 & \alpha_2{e^{-i\eta_1}} & -\alpha_1{e^{-i\eta_2}} & -\alpha_0{e^{-i\eta_3}} &
\alpha_7{e^{-i\eta_4}}& \alpha_6{e^{-i\eta_5}}& -\alpha_5{e^{-i\eta_6}}& -\alpha_4{e^{-i\eta_7}}\\
\alpha_4 & -\alpha_5{e^{-i\eta_1}} & \alpha_6{e^{-i\eta_2}} & -\alpha_7{e^{-i\eta_3}} &
-\alpha_0{e^{-i\eta_4}}& \alpha_1{e^{-i\eta_5}}& -\alpha_2{e^{-i\eta_6}}& \alpha_3{e^{-i\eta_7}}\\
\alpha_5 & \alpha_4{e^{-i\eta_1}} & -\alpha_7{e^{-i\eta_2}} & -\alpha_6{e^{-i\eta_3}} &
-\alpha_1{e^{-i\eta_4}}& -\alpha_0{e^{-i\eta_5}}& \alpha_3{e^{-i\eta_6}}& \alpha_2{e^{-i\eta_7}}\\
\alpha_6 & -\alpha_7{e^{-i\eta_1}} & -\alpha_4{e^{-i\eta_2}} & \alpha_5{e^{-i\eta_3}} &
\alpha_2{e^{-i\eta_4}}& -\alpha_3{e^{-i\eta_5}}& -\alpha_0{e^{-i\eta_6}}& \alpha_1{e^{-i\eta_7}}\\
\alpha_7 & \alpha_6{e^{-i\eta_1}} & \alpha_5{e^{-i\eta_2}} & \alpha_4{e^{-i\eta_3}} &
-\alpha_3{e^{-i\eta_4}}& -\alpha_2{e^{-i\eta_5}}&- \alpha_1{e^{-i\eta_6}}& -\alpha_0{e^{-i\eta_7}}\\
\end{array} \! \! \! \right) \! \! .
\end{eqnarray}
Thus, the total systemic state, encompassing the quantum channels, reads as:
\begin{eqnarray}
|\Phi\rangle &=&|{\phi}_{1}\rangle_{123}\otimes|\phi_{2}\rangle_{456}\otimes|\phi_{3}\rangle_{789}\nonumber \\
&=&\sum_{i,j,k}^{0,1}|{{M}}_{ijk}\rangle_{147}\otimes|{R}_{ijk}\rangle_{235689} \nonumber \\
&=&|{M}_{000}\rangle_{147}(\alpha_0x_0x_1x_2|000000\rangle+\alpha_1e^{i\eta_1}x_0x_1y_2|000011\rangle
+\alpha_2e^{i\eta_2}x_0y_1x_2|001100\rangle \nonumber \\
&&+\alpha_3e^{i\eta_3}x_0y_1y_2|001111\rangle +\alpha_4e^{i\eta_4}y_0x_1x_2|110000\rangle+\alpha_5e^{i\eta_5}y_0x_1y_2|110011\rangle \nonumber \\
&&+\alpha_6e^{i\eta_6}y_0y_1x_2|111100\rangle+\alpha_7e^{i\eta_7}y_0y_1y_2|111111\rangle)_{235689} \nonumber \\
&&+|{M}_{001}\rangle_{147}(\alpha_1x_0x_1x_2|000000\rangle-\alpha_0e^{i\eta_1}x_0x_1y_2|000011\rangle+\alpha_3e^{i\eta_2}x_0y_1x_2|001100\rangle \nonumber \\
&&-\alpha_2e^{i\eta_3}x_0y_1y_2|001111\rangle +\alpha_5e^{i\eta_4}y_0x_1x_2|110000\rangle-\alpha_4e^{i\eta_5}y_0x_1y_2|110011\rangle \nonumber \\
&&+\alpha_7e^{i\eta_6}y_0y_1x_2|111100\rangle-\alpha_6e^{i\eta_7}y_0y_1y_2|111111\rangle)_{235689} \nonumber \\
&&+|{M}_{010}\rangle_{147}(\alpha_2x_0x_1x_2|000000\rangle-\alpha_3e^{i\eta_1}x_0x_1y_2|000011\rangle
-\alpha_0e^{i\eta_2}x_0y_1x_2|001100\rangle \nonumber \\
&&+\alpha_1e^{i\eta_3}x_0y_1y_2|001111\rangle-\alpha_6e^{i\eta_4}y_0x_1x_2|110000\rangle+\alpha_7e^{i\eta_5}y_0x_1y_2|110011\rangle \nonumber \\
&&+\alpha_4e^{i\eta_6}y_0y_1x_2|111100\rangle-\alpha_5e^{i\eta_7}y_0y_1y_2|111111\rangle)_{235689} \nonumber \\
&&+|{M}_{011}\rangle_{147}(\alpha_3x_0x_1x_2|000000\rangle+\alpha_2e^{i\eta_1}x_0x_1y_2|000011\rangle
-\alpha_1e^{i\eta_2}x_0y_1x_2|001100\rangle \nonumber \\
&&-\alpha_0e^{i\eta_3}x_0y_1y_2|001111\rangle + \alpha_7e^{i\eta_4}y_0x_1x_2|110000\rangle+\alpha_6e^{i\eta_5}y_0x_1y_2|110011\rangle \nonumber \\
&&-\alpha_5e^{i\eta_6}y_0y_1x_2|111100\rangle-\alpha_4e^{i\eta_7}y_0y_1y_2|111111\rangle)_{235689} \nonumber \\
&&+|{M}_{100}\rangle_{147}(\alpha_4x_0x_1x_2|000000\rangle-\alpha_5e^{i\eta_1}x_0x_1y_2|000011\rangle
+\alpha_6e^{i\eta_2}x_0y_1x_2|001100\rangle \nonumber \\
&&-\alpha_7e^{i\eta_3}x_0y_1y_2|001111\rangle -\alpha_0e^{i\eta_4}y_0x_1x_2|110000\rangle+\alpha_1e^{i\eta_5}y_0x_1y_2|110011\rangle \nonumber \\
&& -\alpha_2e^{i\eta_6}y_0y_1x_2|111100\rangle+\alpha_3e^{i\eta_7}y_0y_1y_2|111111\rangle)_{235689} \nonumber \\
&&+|{M}_{101}\rangle_{147}(\alpha_5x_0x_1x_2|000000\rangle+\alpha_4e^{i\eta_1}x_0x_1y_2|000011\rangle-\alpha_7e^{i\eta_2}x_0y_1x_2|001100\rangle \nonumber \\
&&-\alpha_6e^{i\eta_3}x_0y_1y_2|001111\rangle - \alpha_1e^{i\eta_4}y_0x_1x_2|110000\rangle-\alpha_0e^{i\eta_5}y_0x_1y_2|110011\rangle \nonumber \\
&&+\alpha_3e^{i\eta_6}y_0y_1x_2|111100\rangle+\alpha_2e^{i\eta_7}y_0y_1y_2|111111\rangle)_{235689} \nonumber \\
&&+|{M}_{110}\rangle_{147}(\alpha_6x_0x_1x_2|000000\rangle-\alpha_7e^{i\eta_1}x_0x_1y_2|000011\rangle
-\alpha_4e^{i\eta_2}x_0y_1x_2|001100\rangle \nonumber \\
&&+\alpha_5e^{i\eta_3}x_0y_1y_2|001111\rangle +\alpha_2e^{i\eta_4}y_0x_1x_2|110000\rangle-\alpha_3e^{i\eta_5}y_0x_1y_2|110011\rangle \nonumber \\
&& -\alpha_0e^{i\eta_6}y_0y_1x_2|111100\rangle+\alpha_1e^{i\eta_7}y_0y_1y_2|111111\rangle)_{235689} \nonumber \\
&&+|{M}_{111}\rangle_{147}(\alpha_7x_0x_1x_2|000000\rangle+\alpha_6e^{i\eta_1}x_0x_1y_2|000011\rangle
+\alpha_5e^{i\eta_2}x_0y_1x_2|001100\rangle \nonumber \\
&&+\alpha_4e^{i\eta_3}x_0y_1y_2|001111\rangle - \alpha_3e^{i\eta_4}y_0x_1x_2|110000\rangle-\alpha_2e^{i\eta_5}y_0x_1y_2|110011\rangle \nonumber \\
&&-\alpha_1e^{i\eta_6}y_0y_1x_2|111100\rangle-\alpha_0e^{i\eta_7}y_0y_1y_2|111111\rangle)_{235689}.
\end{eqnarray}
Within the above, the states $|{R}_{ijk}\rangle$ are non-normalized; ${{N}}_{ijk}$ are the normalized coefficients associated with the states $|{R}_{ijk}\rangle$ and the non-normalized state $|{R}_{ijk}\rangle_{235689}\equiv{_{147}}\langle{M}_{ijk}|\Phi\rangle$ can be obtained with a probability of $ ( 1/{{N}_{ijk}} )^2$.

{\bf Step 2:} In accordance with the measurement outcome $|{M}_{ijk}\rangle$, Alice makes an appropriate triplet joint unitary operation, $\hat{\cal U}_{258}^{ijk}$, on her remaining three qubits: 2, 5 and 8. This operation is performed under the ordering basis:
$\{|000\rangle$, $|001\rangle$, $|010\rangle$, $|011\rangle$, $|100\rangle$, $|101\rangle$, $|110\rangle$, $|111\rangle\}$. To be explicit, $\hat{\cal U}_{258}^{ijk}$ is an $8\times8$ matrix and takes one of the following forms:
\begin{eqnarray}
\hat{\cal{U}}_{258}^{000} \! \! &=& \! \! {\rm diag}(1, 1, 1, 1, 1, 1, 1, 1)=\mathbb{I}_{8\times8}, \\
\hat{\cal{U}}_{258}^{001} \! \! &=& \! \! {\rm diag}(e^{i \theta_1} \! , -e^{-i \theta_1} \! , e^{i( \theta_3 - \theta_2 )} \! , -e^{i( \theta_2 - \theta_3 )} \! , e^{i( \theta_5 - \theta_4 )} \! , -e^{i( \theta_4 - \theta_5 )} \! , e^{i( \theta_7 - \theta_6 )} \! , -e^{i( \theta_6 - \theta_7 )} \! ) , \\
\hat{\cal{U}}_{258}^{010} \! \! &=& \! \! {\rm diag}(e^{i\theta_2} \! , -e^{i(\theta_3-\theta_1)} \! , -e^{-i\theta_2} \! , e^{i(\theta_1-\theta_3)} \! , -e^{i(\theta_6-\theta_4)} \! , e^{i(\theta_7-\theta_5)} \! , e^{i(\theta_4-\theta_6)} \! , -e^{i(\theta_5-\theta_7)} \! ) , \\
\hat{\cal{U}}_{258}^{011} \! \! &=& \! \! {\rm diag}(e^{i\theta_3} \! , e^{i(\theta_2-\theta_1)} \! , -e^{i(\theta_1-\theta_2)} \! , -e^{-i\theta_3} \! , e^{i(\theta_7-\theta_4)} \! , e^{i(\theta_6-\theta_5)} \! , -e^{i(\theta_5-\theta_6)} \! , -e^{i(\theta_4-\theta_7)} \! ) , \\
\hat{\cal{U}}_{258}^{100} \! \! &=& \! \! {\rm diag}(e^{i\theta_4} \! , -e^{i(\theta_5-\theta_1)} \! , e^{i(\theta_6-\theta_2)} \! , -e^{i(\theta_7-\theta_3)} \! , -e^{i-\theta_4} \! , e^{i(\theta_1-\theta_5)} \! , -e^{i(\theta_2-\theta_6)} \! , e^{i(\theta_3-\theta_7)} \! ) , \\
\hat{\cal{U}}_{258}^{101} \! \! &=& \! \! {\rm diag}(e^{i\theta_5} \! , e^{i(\theta_4-\theta_1)} \! , -e^{i(\theta_7-\theta_2)} \! , -e^{i(\theta_6-\theta_3)} \! , -e^{i(\theta_1-\theta_4)} \! , -e^{-i\theta_5} \! , e^{i(\theta_3-\theta_6)} \! , e^{i(\theta_2-\theta_7)} \! ) , \\
\hat{\cal{U}}_{258}^{110} \! \! &=& \! \! {\rm diag}(e^{i\theta_6} \! , -e^{i(\theta_7-\theta_1)} \! , -e^{i(\theta_4-\theta_2)} \! , e^{i(\theta_5-\theta_3)} \! , e^{i(\theta_2-\theta_4)} \! , -e^{i(\theta_3-\theta_5)} \! , -e^{-i\theta_6} \! , e^{i(\theta_1-\theta_7)} \! ) , \\
\hat{\cal{U}}_{258}^{111} \! \! &=& \! \! {\rm diag}(e^{i\theta_7} \! , e^{i(\theta_6-\theta_1)} \! , e^{i(\theta_5-\theta_2)} \! , e^{i(\theta_4-\theta_3)} \! , -e^{i(\theta_3-\theta_4)} \! , -e^{i(\theta_2-\theta_5)} \! , -e^{i(\theta_1-\theta_6)} \! , -e^{-i\theta_7} \! ) .
\end{eqnarray}

{\bf Step 3:} Next, Alice performs a measurement on her Qubits 2, 5 and 8 under the set of complete orthogonal basis vector $\{|\pm\rangle\}$ and broadcasts the measurement outcome via a classical channel (\textit{i.e}., sending some cbits). Again, all of the authorized anticipators make an agreement in advance that cbits $'ijk'$ correspond to the outcome $|{M}_{ijk}\rangle_{147}$ ($(i_1 i_2 i_3)$ within {Section 2} of main text) and cbits $'rst'$ to the measuring outcome of Qubits 2, 5 and 8 ($(j_1 j_2 j_3)$ within {Section 2} of main text), respectively.

{\bf Step 4:} After receiving Alice's messages, Bob introduces one auxiliary qubit, $A$, with an initial state of $|0\rangle$. Bob then makes quadruplet collective unitary transformation, $\hat{\cal U}_{369A}$, on Qubits 3, 6, 9 and $A$ under a set of ordering basis vectors: $\{|0000\rangle_{369A},$ $|0010\rangle_{369A},$ $|0100\rangle_{369A},$ $|0110\rangle_{369A},$ $|1000\rangle_{369A},$ $|1010\rangle_{369A},$ $|1100\rangle_{369A},$ $|1110\rangle_{369A},$
$\{|0001\rangle_{369A},$ $|0011\rangle_{369A},$ $|0101\rangle_{369A},$ $|0111\rangle_{369A},$ $|1001\rangle_{369A},$ $|1011\rangle_{369A},$ $|1101\rangle_{369A},$ $|1111\rangle_{369A}$. The form of this transformation operator is:
\begin{eqnarray}
\hat{\cal{U}}_{369A}=\left(
\begin{array}{cc}
{\hat{\cal{H}}}& {\hat{\cal{G}}}\\
{\hat{\cal{G}}} & -{\hat{\cal{H}}}\\
\end{array}
\right)_{16\times16},
\end{eqnarray}
where ${\hat{\cal{H}}}$ and ${\hat{\cal{G}}}$ are both $8\times8$ matrices. Explicitly, these matrices are given by:
\begin{equation}
{\hat{\cal{H}}}={\rm diag}\left(1,\frac{x_2}{y_2},\frac{x_1}{y_1},\frac{x_1x_2}{y_1y_2},\frac{x_0}{y_0},\frac{x_0x_2}{y_0y_2},\frac{x_0x_1}{y_0y_1},\frac{x_0x_1x_2}{y_0y_1y_2}\right)
\end{equation}
and:
\begin{equation}
\begin{split}
{\hat{\cal{G}}}={\rm diag} &\bigg(0,\sqrt{1-(\frac{x_2}{y_2})^2},\sqrt{1-(\frac{x_1}{y_1})^2},\sqrt{1-(\frac{x_1x_2}{y_1y_2})^2},\sqrt{1-(\frac{x_0}{y_0})^2},\\&\sqrt{1-(\frac{x_0x_2}{y_0y_2})^2}, \sqrt{1-(\frac{x_0x_1}{y_0y_1})^2},\sqrt{1-(\frac{x_0x_1x_2}{y_0y_1y_2})^2}~\bigg).
\end{split}
\end{equation}

Next, Bob measures his auxiliary qubit, $A$, under the set of measuring basis vectors:
$\{|0\rangle,|1\rangle\}$. If state $|1\rangle$ is measured, his remaining
qubits will collapse into the trivial state, leading to the failure of the RSP. Otherwise,
$|0\rangle$ is obtained, and the procedure shall continue forward to the final step.

{\bf Step 5:} Finally, Bob operates with an appropriate unitary transformation, $\hat{{\cal U}}_{369}^{ijkrst}$
(see Table \ref{apptab} for details), on Qubits 3, 6 and 9.

\begin{figure}
\centering
\includegraphics[scale=1]{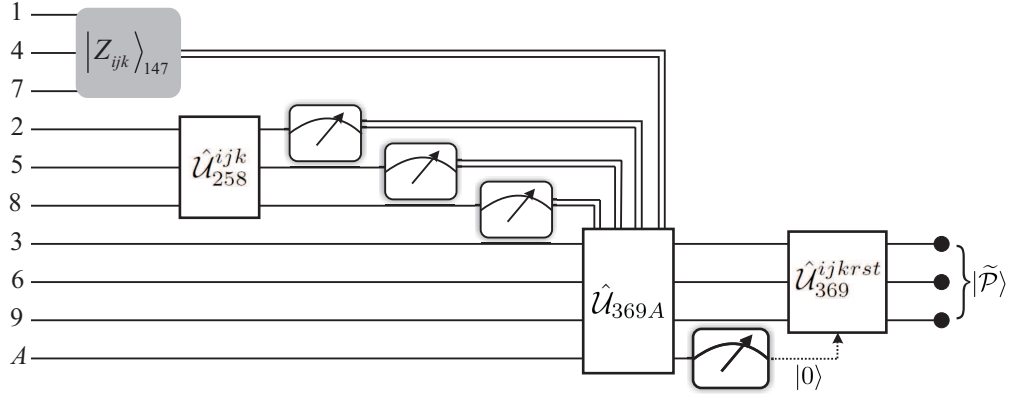}
\caption{Quantum circuit for implementing RSP of arbitrary three-qubit entangled states. $|{M}_{ijk}\rangle_{147}$ denotes a three-qubit projective measurement on
Qubits 1, 4 and 7 under
a set of complete orthogonal basis vectors $\{|{M}_{ijk}\rangle_{147}\}$; $\hat{\cal U}_{258}^{ijk}$ denotes Alice's
appropriate triplet collective unitary transformation on triplet (2,5,8);
$\hat{\cal U}_{369A}$ denotes Bob's collective four-qubit unitary transformation on his Qubits 3, 6, 9 and $A$ and $\hat{\cal U}_{369}^{ijkrst}$ denotes Bob's
appropriate single-qubit unitary transformations on his Qubits 3, 6 and 9.}
\label{fig:2}
\end{figure}

For clarity, the quantum circuit for this RSP scheme is provided as
Figure~\ref{fig:2}.

\begin{table}
\caption{$ijkrst$ denotes the CIC corresponding to measurement outcomes from the sender; $\hat{\cal{U}}_{369}^{ijkrst}$
denotes the unitary transformation that Bob needs to perform on Qubits 3, 6 and 9 to recover the desired state, $ | \cal{P} \rangle $.}
\centering
{\small
\begin{tabular}{|l|l|l|l||l|l|l|l|}
\hline $ijkrst$ & $\hat{\cal{U}}_{369}^{ijkrst}$ & $ijkrst$ & $\hat{\cal{U}}_{369}^{ijkrst}$ & $ijkrst$ &
$\hat{\cal{U}}_{369}^{ijkrst}$& $ijkrst$ & $\hat{\cal{U}}_{369}^{ijkrst}$\\ \hline
$000000$ & $I_{3}I_{6}I_{9}$ & $010000$ & $I_{3}\sigma^x_{6}I_{9}$ & $100000$ & $\sigma^x_{3}I_{6}I_{9}$ &$110000$ & $\sigma^x_3\sigma^x_6I_9$ \\
$000001$ & $I_{3}I_{6}\sigma_9^z$ & $010001$ & $I_{3}\sigma^x_{6}\sigma^z_{9}$ & $100001$ & $\sigma^x_{3}I_{6}\sigma^z_{9}$ &$110001$ & $\sigma^x_3\sigma^x_6\sigma^z_9$ \\
$000010$ & $I_{3}\sigma_6^zI_9$ & $010010$ & $I_{3}\sigma^x_{6}\sigma^z_{6}I_9$ & $100010$ & $\sigma^x_{3}\sigma^z_{6}I_{9}$ &$110010$ & $\sigma^x_3\sigma^x_6\sigma^z_6I_9$ \\
$000011$ & $I_{3}\sigma^z_{6}\sigma_9^z$ & $010011$ & $I_{3}\sigma^x_{6}\sigma^z_{6}\sigma^z_9$ & $100011$ & $\sigma^x_{3}\sigma^z_{6}\sigma^z_{9}$ &$110011$ & $\sigma^x_3\sigma^x_6\sigma^z_6\sigma^z_9$ \\
$000100$ & $\sigma^z_{3}I_{6}I_9$ & $010100$ & $\sigma^z_{3}\sigma^x_{6}I_9$ & $100100$ & $\sigma^x_{3}\sigma^z_{3}I_6I_{9}$ &$110100$ & $\sigma^x_3\sigma^z_3\sigma^x_6I_9$ \\
$000101$ & $\sigma^z_{3}I_{6}\sigma_9^z$ & $010101$ & $\sigma^z_{3}\sigma^x_{6}\sigma^z_9$& $100101$ & $\sigma^x_{3}\sigma^z_{3}I_6\sigma^z_{9}$ &$110101$ & $\sigma^x_3\sigma^z_3\sigma^x_6\sigma^z_9$ \\
$000110$ & $\sigma^z_{3}\sigma^z_{6}I_9$ & $010110$ & $\sigma^z_{3}\sigma^x_{6}\sigma^z_6I_9$ & $100110$ & $\sigma^x_{3}\sigma^z_{3}\sigma^z_6I_{9}$ &$110110$ & $\sigma^x_3\sigma^z_3\sigma^x_6\sigma^z_6I_9$ \\
$000111$ & $\sigma^z_{3}\sigma^z_{6}\sigma_9^z$ & $010111$ & $\sigma^z_{3}\sigma^x_{6}\sigma^z_6\sigma^z_9$ & $100111$ & $\sigma^x_{3}\sigma^z_{3}\sigma^z_6\sigma^z_{9}$ &$110111$ & $\sigma^x_3\sigma^z_3\sigma^x_6\sigma^z_6\sigma^z_9$ \\
$001000$ & $I_3I_6\sigma^x_9$ & $011000$ & $I_{3}\sigma^x_{6}\sigma^x_{9}$ & $101000$ & $\sigma^x_3I_6\sigma^x_9$ &$111000$ & $\sigma^x_3\sigma^x_6\sigma^x_9$ \\
$001001$ & $I_3I_6\sigma^x_9\sigma^z_9$ & $011001$ & $I_{3}\sigma^x_{6}\sigma^x_{9}\sigma^z_9$& $101001$ & $\sigma^x_3I_6\sigma^x_9\sigma^z_9$ &$111001$ & $\sigma^x_3\sigma^x_6\sigma^x_9\sigma^z_9$ \\
$001010$ & $I_3\sigma^z_6\sigma^x_9$ & $011010$ & $I_{3}\sigma^x_{6}\sigma^z_6\sigma^x_{9}$ & $101010$ & $\sigma^x_3\sigma^z_6\sigma^x_9$ &$111010$ & $\sigma^x_3\sigma^x_6\sigma^z_6\sigma^x_9$ \\
$001011$ & $I_3\sigma^z_6\sigma^x_9\sigma^z_9$ & $011011$ & $I_{3}\sigma^x_{6}\sigma^z_6\sigma^x_{9}\sigma^z_9$ & $101011$ & $\sigma^x_3\sigma^z_6\sigma^x_9\sigma^z_9$ &$111011$ & $\sigma^x_3\sigma^x_6\sigma^z_6\sigma^x_9\sigma^z_9$ \\
$001100$ & $\sigma^z_3I_6\sigma^x_9$ & $011100$ & $\sigma^z_{3}\sigma^x_{6}\sigma^x_{9}$ & $101100$ & $\sigma^x_3\sigma^z_3I_6\sigma^x_9$ &$111100$ & $\sigma^x_3\sigma^z_3\sigma^x_6\sigma^x_9$ \\
$001101$ & $\sigma^z_3I_6\sigma^x_9\sigma^z_9$ & $011101$ & $\sigma^z_{3}\sigma^x_{6}\sigma^x_{9}\sigma^z_9$& $101101$ & $\sigma^x_3\sigma^z_3I_6\sigma^x_9\sigma^z_9$ &$111101$ & $\sigma^x_3\sigma^z_3\sigma^x_6\sigma^x_9\sigma^z_9$ \\
$001110$ & $\sigma^z_3\sigma^z_6\sigma^x_9$ & $011110$ & $\sigma^z_{3}\sigma^x_{6}\sigma^z_{9}\sigma^x_9$ & $101110$ & $\sigma^x_3\sigma^z_3\sigma^z_6\sigma^x_9$ &$111110$ & $\sigma^x_3\sigma^z_3\sigma^x_6\sigma^z_6\sigma^x_9$ \\
$001111$ & $\sigma^z_3\sigma^z_6\sigma^x_9\sigma^z_9$ & $011111$ & $\sigma^z_{3}\sigma^x_{6}\sigma^z_6\sigma^x_{9}\sigma^z_9$ & $101111$ & $\sigma^x_3\sigma^z_3\sigma^z_6\sigma^x_9\sigma^z_z$ &$111111$ & $\sigma^x_3\sigma^z_3\sigma^x_6\sigma^z_6\sigma^x_9\sigma^z_9$ \\
\hline
\end{tabular}}
\label{apptab}
\end{table}

\section{Three-Qubit Entangled State RSP for $(i,j,k,r,s,t)=(0,0,1,0,0,1)$}
\label{General-2}

Above, we have shown that RSP for an arbitrary three-qubit entangled state can be faithfully performed with a certain success probability. For clarity, here we will take the case of $(i,j,k,r,s,t)=(0,0,1,0,0,1)$ as an example. That is, the state $|{M}_{001}\rangle_{147}$ is detected by Alice at the beginning. Thus, the remaining qubits will be converted into:
\begin{equation} \begin{split}
|R_{001}\rangle_{235689}=&(\alpha_1x_0x_1x_2|000000\rangle \alpha_0e^{i\eta_1}x_0x_1y_2|000011\rangle
+\alpha_3e^{i\eta_2}x_0y_1x_2|001100\rangle \\
&-\alpha_2e^{i\eta_3}x_0y_1y_2|001111\rangle +\alpha_5e^{i\eta_4}y_0x_1x_2|110000\rangle-\alpha_4e^{i\eta_5}y_0x_1y_2|110011\rangle \\
&+\alpha_7e^{i\eta_6}y_0y_1x_2|111100\rangle-\alpha_6e^{i\eta_7}y_0y_1y_2|111111\rangle)_{235689}.
\end{split}
\end{equation}

Later, Alice makes the operation $\hat{\cal U}_{258}^{001}$ on her remaining Qubits 2, 5 and 8. As a consequence, the above state will evolve into:
\begin{equation}
\begin{split}
&{N}_{001}
(\alpha_1e^{i\eta_1}x_0x_1x_2|000000\rangle+\alpha_0x_0x_1y_2|000011\rangle
+\alpha_3e^{i\eta_3}x_0y_1x_2|001100\rangle \\
&+\alpha_2e^{i\eta_2}x_0y_1y_2|001111\rangle+\alpha_5e^{i\eta_5}y_0x_1x_2|110000\rangle+\alpha_4e^{i\eta_4}y_0x_1y_2|110011\rangle \\
&+\alpha_7e^{i\eta_7}y_0y_1x_2|111100\rangle+\alpha_6e^{i\eta_6}y_0y_1y_2|111111\rangle)_{235689}.
\label{equation 1}
\end{split}
\end{equation}
Within the above, the normalization parameter is: ${N}_{001}\equiv(|\alpha_1x_0x_1x_2|^2+|\alpha_0x_0x_1y_2|^2+
|\alpha_3x_0y_1x_2|^2+|\alpha_2x_0y_1y_2|^2+|\alpha_5y_0x_1x_2|^2+|\alpha_4y_0x_1y_2|^2
+|\alpha_7y_0y_1x_2|^2+|\alpha_6y_0y_1y_2|^2)^{-\frac12}$. Incidentally, the state given in Equation (\ref{equation 1}) can be rewritten as:
\begin{eqnarray}
|R_{001}\rangle\!\!&=&\sum_{r,s,t}^{0,1}|\varphi_{rst}\rangle_{258}|\psi_{rst}\rangle_{369}\nonumber \\ \nonumber
&=&\frac{{N}_{001}}{2\sqrt2}[|+++\rangle_{258}(\alpha_1e^{i\eta_1}x_0x_1x_2|000\rangle+\alpha_0x_0x_1y_2|001\rangle +\alpha_3e^{i\eta_3}x_0y_1x_2|010\rangle \\ \nonumber
&&+\alpha_2e^{i\eta_2}x_0y_1y_2|011\rangle+\alpha_5e^{i\eta_5}y_0x_1x_2|100\rangle+\alpha_4e^{i\eta_4}y_0x_1y_2|101\rangle \\ \nonumber
&&+\alpha_7e^{i\eta_7}y_0y_1x_2|110\rangle+\alpha_6e^{i\eta_6}y_0y_1y_2|111\rangle)_{369}\\ \nonumber
&&+|++-\rangle_{258}(\alpha_1e^{i\eta_1}x_0x_1x_2|000\rangle-\alpha_0x_0x_1y_2|001\rangle +\alpha_3e^{i\eta_3}x_0y_1x_2|010\rangle \\ \nonumber
&&-\alpha_2e^{i\eta_2}x_0y_1y_2|011\rangle +\alpha_5e^{i\eta_5}y_0x_1x_2|100\rangle-\alpha_4e^{i\eta_4}y_0x_1y_2|101\rangle \\ \nonumber
&&+\alpha_7e^{i\eta_7}y_0y_1x_2|110\rangle-\alpha_6e^{i\eta_6}y_0y_1y_2|111\rangle)_{369}\\ \nonumber
&&+|+-+\rangle_{258}(\alpha_1e^{i\eta_1}x_0x_1x_2|000\rangle+\alpha_0x_0x_1y_2|001\rangle -\alpha_3e^{i\eta_3}x_0y_1x_2|010\rangle \\ \nonumber
&&-\alpha_2e^{i\eta_2}x_0y_1y_2|011\rangle +\alpha_5e^{i\eta_5}y_0x_1x_2|100\rangle+\alpha_4e^{i\eta_4}y_0x_1y_2|101\rangle \\ \nonumber
&&-\alpha_7e^{i\eta_7}y_0y_1x_2|110\rangle-\alpha_6e^{i\eta_6}y_0y_1y_2|111\rangle)_{369}\\ \nonumber
&&+|+--\rangle_{258}(\alpha_1e^{i\eta_1}x_0x_1x_2|000\rangle-\alpha_0x_0x_1y_2|001\rangle -\alpha_3e^{i\eta_3}x_0y_1x_2|010\rangle \\ \nonumber
&&+\alpha_2e^{i\eta_2}x_0y_1y_2|011\rangle+\alpha_5e^{i\eta_5}y_0x_1x_2|100\rangle-\alpha_4e^{i\eta_4}y_0x_1y_2|101\rangle \\ \nonumber
&&-\alpha_7e^{i\eta_7}y_0y_1x_2|110\rangle+\alpha_6e^{i\eta_6}y_0y_1y_2|111\rangle)_{369}\\ \nonumber
&&+|-++\rangle_{258}(\alpha_1e^{i\eta_1}x_0x_1x_2|000\rangle+\alpha_0x_0x_1y_2|001\rangle +\alpha_3e^{i\eta_3}x_0y_1x_2|010\rangle \\ \nonumber
&&+\alpha_2e^{i\eta_2}x_0y_1y_2|011\rangle -\alpha_5e^{i\eta_5}y_0x_1x_2|100\rangle-\alpha_4e^{i\eta_4}y_0x_1y_2|101\rangle \\ \nonumber
&&-\alpha_7e^{i\eta_7}y_0y_1x_2|110\rangle-\alpha_6e^{i\eta_6}y_0y_1y_2|111\rangle)_{369}\\ \nonumber
&&+|-+-\rangle_{258}(\alpha_1e^{i\eta_1}x_0x_1x_2|000\rangle-\alpha_0x_0x_1y_2|001\rangle +\alpha_3e^{i\eta_3}x_0y_1x_2|010\rangle \\ \nonumber
&&-\alpha_2e^{i\eta_2}x_0y_1y_2|011\rangle -\alpha_5e^{i\eta_5}y_0x_1x_2|100\rangle+\alpha_4e^{i\eta_4}y_0x_1y_2|101\rangle \\ \nonumber
&&-\alpha_7e^{i\eta_7}y_0y_1x_2|110\rangle+\alpha_6e^{i\eta_6}y_0y_1y_2|111\rangle)_{369}\\ \nonumber
&&+|--+\rangle_{258}(\alpha_1e^{i\eta_1}x_0x_1x_2|000\rangle+\alpha_0x_0x_1y_2|001\rangle-\alpha_3e^{i\eta_3}x_0y_1x_2|010\rangle \\ \nonumber
&&-\alpha_2e^{i\eta_2}x_0y_1y_2|011\rangle -\alpha_5e^{i\eta_5}y_0x_1x_2|100\rangle-\alpha_4e^{i\eta_4}y_0x_1y_2|101\rangle \\ \nonumber
&&+\alpha_7e^{i\eta_7}y_0y_1x_2|110\rangle+\alpha_6e^{i\eta_6}y_0y_1y_2|111\rangle)_{369}\\ \nonumber
&&+|---\rangle_{258}(\alpha_1e^{i\eta_1}x_0x_1x_2|000\rangle-\alpha_0x_0x_1y_2|001\rangle-\alpha_3e^{i\eta_3}x_0y_1x_2|010\rangle \\ \nonumber
&&+\alpha_2e^{i\eta_2}x_0y_1y_2|011\rangle-\alpha_5e^{i\eta_5}y_0x_1x_2|100\rangle+\alpha_4e^{i\eta_4}y_0x_1y_2|101\rangle \\ \nonumber
&&+\alpha_7e^{i\eta_7}y_0y_1x_2|110\rangle-\alpha_6e^{i\eta_6}y_0y_1y_2|111\rangle)_{369}].
\end{eqnarray}

Accordingly, Alice measures Qubits 2, 5 and 8 under the basis vectors $\{|\pm\rangle\}$. Letting the outcome be $|+\rangle_{2}|+\rangle_5|-\rangle_{8}$, Alice broadcasts this outcome to Bob via the classical message '001'. The subsystem state will then be:
\begin{equation}
\begin{split}
|\psi_{001}\rangle_{369}=&N_{001}(\alpha_1e^{i\eta_1}x_0x_1x_2|000\rangle-\alpha_0x_0x_1y_2|001\rangle
+\alpha_3e^{i\eta_3}x_0y_1x_2|010\rangle \\
&-\alpha_2e^{i\eta_2}x_0y_1y_2|011\rangle+\alpha_5e^{i\eta_5}y_0x_1x_2|100\rangle-\alpha_4e^{i\eta_4}y_0x_1y_2|101\rangle \\
&+\alpha_7e^{i\eta_7}y_0y_1x_2|110\rangle-\alpha_6e^{i\eta_6}y_0y_1y_2|111\rangle)_{369}.
\end{split}
\end{equation}

Bob then introduces the auxiliary qubit, $A$, with an initial state of $|0\rangle$. He may now implement a local quadruplet collective unitary transformation, $\hat{\cal U}_{369A}$, on Qubits 3, 6, 9 and $A$. Thus, Bob's system will become:

\begin{equation}
\begin{split}
&{N}_{001}[x_0x_1x_2(\alpha_1{e^{i\eta_1}}|000\rangle-\alpha_0|001\rangle
+\alpha_3{e}^{i\eta_3}|010\rangle-\alpha_2{e}^{i\eta_2}|011\rangle+\alpha_5{e}^{i\eta_5}|100\rangle\\
&-\alpha_4{e}^{i\eta_4}|101\rangle
+\alpha_7{e}^{i\eta_7}|110\rangle-\alpha_6{e}^{i\eta_6}|111\rangle)_{369}\otimes|0\rangle_{A}+(-\alpha_0x_0x_1y_2\sqrt{1-(\frac{x_2}{y_2})^2}|001\rangle\\
&-\alpha_2e^{i\eta_2}x_0y_1y_2\sqrt{1-(\frac{x_1x_2}{y_1y_2})^2}|011\rangle
+\alpha_3e^{i\eta_3}x_0y_1x_2\sqrt{1-(\frac{x_1}{y_1})^2}|010\rangle\\
&-\alpha_4e^{i\eta_4}y_0x_1y_2\sqrt{1-(\frac{x_0x_2}{y_0y_2})^2}|101\rangle+\alpha_5e^{i\eta_5}y_0x_1x_2\sqrt{1-(\frac{x_0}{y_0})^2}|100\rangle\\
&-\alpha_6e^{i\eta_6}y_0y_1y_2\sqrt{1-(\frac{x_0x_1x_2}{y_0y_1y_2})^2}|111\rangle
+\alpha_7e^{i\eta_7}y_0y_1x_2\sqrt{1-(\frac{x_0x_1}{y_0y_1})^2}|110\rangle)_{369}\otimes|1\rangle_{A}].
\end{split}
\end{equation}

Subsequently, he makes a single-qubit projective measurement on qubit $A$ under basis vectors
$\{|0\rangle,|1\rangle\}$. If $|1\rangle_{A}$ is measured, his remaining qubits will collapse into the trivial state, and the RSP fails. If $|0\rangle_{A}$ is measured, the remaining qubits will transform into the state: $(\alpha_1{e^{i\eta_1}}|000\rangle-\alpha_0|001\rangle
+\alpha_3{e}^{i\eta_3}|010\rangle-\alpha_2{e}^{i\eta_2}|011\rangle+\alpha_5{e}^{i\eta_5}|100\rangle
-\alpha_4{e}^{i\eta_4}|101\rangle
+\alpha_7{e}^{i\eta_7}|110\rangle-\alpha_6{e}^{i\eta_6}|111\rangle)_{369}
\equiv{(\hat{\cal U}_{369}^{001001})}^{\dag}| {\cal P }\rangle$. This may readily allow Bob to redeem the desired state after the operation: $\hat{\cal U}_{369}^{001001}=I_3I_6\sigma_9^x\sigma_9^z$.

Of course, Alice's outcome may be one of the remaining seven states: $|{M}_{000}\rangle$, $|{M}_{010}\rangle$, $|{M}_{011}\rangle$, $|{M}_{100}\rangle$, $|{M}_{101}\rangle$, $|{M}_{110}\rangle$ and $|{M}_{111}\rangle$. Therefore, the desired state can be faithfully recovered at Bob's location with certainty by similar analysis methods as those above.
\\

\noindent {\bf Acknowledgements} This work was supported by the program for the National Natural Science
Foundation of China (Grant Nos. 11247256, 11074002 and 61275119), the fund of Anhui Provincial Natural Science Foundation (Grant No. 1508085QF139),
the fund of China Scholarship Council, and the fund from National Laboratory for Infrared Physics (Grant No. M201307).

\end{document}